\documentclass[useAMS,usenatbib]{mn2e}
\usepackage{graphicx,rotate,url,mathptmx,lscape,times,amssymb,color}
\def\gtsim{\mathrel{\hbox{\rlap{\hbox{\lower4pt\hbox{$\sim$}}}\hbox{$>$}}}}
\def\lesssim{\mathrel{\hbox{\rlap{\hbox{\lower4pt\hbox{$\sim$}}}\hbox{$<$}}}}

\def\ga{{\rm\thinspace gauss}}

\def\h0{\hbox{{\rm H}$^0$}}
\DeclareMathAlphabet{\vib}{OML}{cmm}{m}{it}

\title[Improved He~I Emissivities]{Improved He~I Emissivities in the Case B Approximation}

\author[R.L. Porter et al.]
       {\parbox[]{6.0in}
        {R.L. Porter$^{1}$\thanks{E-mail: ryanlporter@gmail.com},
        G.J. Ferland$^{2}$,
        P.J. Storey$^{3}$,
        M.J. Detisch$^{2}$\\
        \footnotesize
        $^1$Department of Physics and Astronomy and Center for Simulational Physics, University of Georgia, Athens, GA 30602, USA\\
        $^2$Dept. of Physics \& Astronomy, University of Kentucky, Lexington, KY 40506, USA\\
        $^3$Dept. of Physics \& Astronomy, University College London, Gower Street, London WC1E 6BT, UK}}
\date{Accepted.
      Received }

\pagerange{\pageref{firstpage}--\pageref{lastpage}}
\pubyear{}

\begin{document}

\maketitle

\label{firstpage}

\begin{abstract}

We update our prior work on the case B collisional-recombination spectrum of He~I to
incorporate \textit{ab initio} photoionisation cross-sections.
This large set of accurate, self-consistent cross-sections represents a
significant improvement in He~I emissivity calculations
because it largely obviates the piecemeal nature that has marked all modern works.
A second, more recent set of \textit{ab initio} cross-sections is also available, 
but we show that those are less consistent with bound-bound transition 
probabilities than our adopted set.
We compare our new effective recombination coefficients with our prior work 
and our new emissivities with those by other researchers,
and we conclude with brief remarks on the effects of the present work on the He~I error budget.
Our calculations cover temperatures $5000 \le T_e \le 25000$~K and densities $10^1 \le n_e \le 10^{14}$~cm$^{-3}$.
Full results are available online.

\end{abstract}


\section{Introduction}
\label{intro}

The high precision required to make a significant measurement of the 
primordial helium abundance presents challenges that are unique in nebular astrophysics.  
The recombination rate coefficients, needed to convert emission line intensities into ionic abundance ratios, 
must have an accuracy better than the precision expected for the derived abundance.  
Usually we hope to measure abundances of most elements to an accuracy of 20 - 30\%.  
The primordial helium problem requires recombination rates accurate to better than a percent, 
presenting unprecedented challenges to the atomic physics of the line formation.
See Brocklehurst (1972) and references therein for early seminal work in the field of 
theoretical He~I emissivities.

We need four types of atomic data to calculate the case B recombination spectrum:
level energies, transition probabilities, photoionization cross-sections, and collision rates.
Porter et al (2009) summarize the contributors to the error budget.
Energy uncertainties have always been negligible compared to other sources, and
accurate transition probabilities have been available since Kono \& Hattori (1984).
Drake (1996) and Drake \& Morton (2007) improved upon the latter still further.

Photoionisation cross-sections and collision rates represent the greatest sources of uncertainties 
in current standard calculations of He~I emissivities 
(Benjamin, Skillman, \& Smits 1999, hereafter BSS99;  Porter et al. 2005, 2007).
For the low-density, extragalactic observations used in primordial helium analyses 
(Peimbert, Luridiana, \& Peimbert 2007; Izotov, Thuan, \& Stasi\'nska 2007;
Aver, Olive, \& Skillman 2010),
photoionisation cross-sections (and, by extension, recombination coefficients)
are the greatest source of uncertainty (Porter et al. 2009).
See Ferland et al. (2010) for a recent review of the errors in He~I emissivities and 
in primordial helium abundances.

In this Letter we update our earlier calculations of case B, 
He~I emissivities (Bauman et al. 2005; Porter et al. 2005, 2007) to include a large set of self-consistent, 
\textit{ab initio} photoionisation cross-sections.
We compare our present results with our previous work, with those by BSS99, 
and with those by Almog \& Netzer (1989).
We present here a subset of our present results that is most applicable to primordial helium research.
A much larger set is available in the electronic edition.
 
\section{Atomic Data}

The first large set ($L \le 4$ and $n \le 25$) of modern, self-consistent photoionisation cross-sections
was calculated by Hummer \& Storey (1998, hereafter HS98).
However, the full set has remained unpublished.
Ercolano \& Storey (2006) used the full set to produce accurate continuous emission spectra.
The current work represents the first use in case B line emissivity calculations.

\begin{figure}
\protect\resizebox{\columnwidth}{!}
{\includegraphics{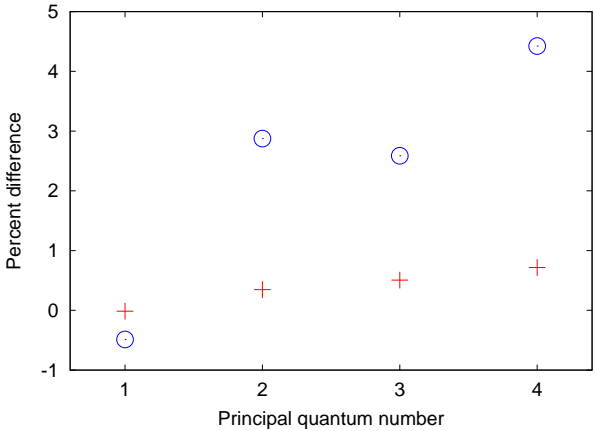}}
\caption{
Percentage differences for threshold photoionisation cross-sections of $ns\,^1\!S$ states relative to ``extrapolated'' Drake (1996) results.  
Nahar (2010, blue circles), HS98 (red crosses).}
\label{fig:compare2}
\end{figure}

Another large set of photoionisation cross-sections was presented recently by Nahar (2010).
Because highly-accurate bound-bound absorption oscillator strengths obtained with Hylleras-type wave functions are available (Drake 1996), 
we can readily apply a simple extrapolation method (Burgess \& Seaton 1960) 
to judge the accuracy of free-bound differential oscillator strengths at the ionization threshold. 
HS98 applied this method to several low-lying levels of He~I and tabulated results in their Table 1.
In Figure~\ref{fig:compare2}, we compare those results with the \textit{ab initio} calculations of $ns\,^1\!S$ levels
from both Nahar and HS98. 
Nahar's values disagree with the extrapolated cross-sections by as much as $4\%$.
The HS98 cross-sections are clearly more consistent with the Drake results than are the Nahar (2010) values. 
The target wave functions used in HS98 were derived
from those of Calvert \& Davison (1971) and were designed to fully
represent the dipole and quadrupole polarizability of the He$^+$
ground state and to include short-range correlation optimised on the
energies of the $ns\,^1\!S$ states, which were known to be particularly
difficult to represent. 
This target provides very accurate results at
energies near threshold, as required for the temperature range and
applications discussed here but is unsuitable at higher energies. 
The target used by Nahar (2010) is extensive and implicitly treats
polarization and correlation effects but it is not optimised for near
threshold calculation or for the $ns\,^1\!S$ states. 
It is, however, applicable over a much larger energy range than that used in HS98.

We use the unpublished cross-sections described by HS98.
For $n\le4$ and $L\le2$, we rescale to agree at threshold to those extrapolated from the Drake oscillator strengths and
listed in table 1 of HS98.
We use the extrapolation procedure described by HS98 to obtain values for $n=5$ and $L\le2$,
but these normalisations have only minor effects on the emissivities of the most important lines.

We follow HS98 and assume that recombination coefficients are fixed relative to hydrogen
for $n > 25$ and $L \le 2$.
All other aspects of our calculations are as described in Bauman et al. (2005).
Briefly, this entails $nLS$-resolved levels for $n \le 100$, a single ``collapsed'' level at $n=101$,
and collisions induced by electrons, protons, and He$^+$. 
The calculations are performed with a development version of the spectral simulation code \textsc{Cloudy} 
(last described by Ferland et al. 1998).
\footnote{The $J$-resolved code described by Bauman et al. (2005) 
has also been updated with the HS98 cross-sections.
That code can be retrieved with a Subversion client from https://svn.nublado.org/bauman/source.}

Following Switzer \& Hirata (2008), 
we also have added radiative bound-bound electric quadrupole transitions (Cann \& Thakkar 2002).
However, these transitions have a negligible effect on the present results;
we mention the change for completeness 
and to note that these transitions may be important for high-Z He-like ions (as in Porter \& Ferland 2007).

Finally, we note that we have compared our existing implementation of 
heavy particle angular momentum changing collisions to the new results by
Vrinceanu, Onofrio, and Sadeghpour (2012) and found excellent agreement.

\section{Results and Discussion}

\begin{figure}
\protect\resizebox{\columnwidth}{!}
{\includegraphics{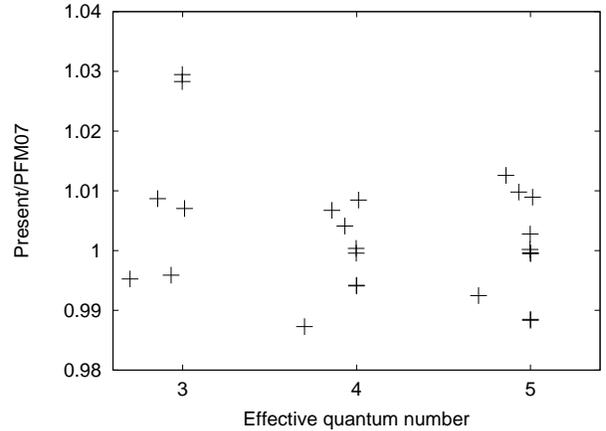}}
\caption{Ratio of present to PFM07 effective recombination coefficients in the low-density limit at $10,000$K.  The three levels with
the largest changes are $3d\,^1\!D$, $3d\,^3\!D$, and $5s\,^1\!S$.  See text.  
The differences will generally be smaller at finite densities.}
\label{fig:alpha}
\end{figure}

We compare effective recombination coefficients in our new treatment 
with those of Porter et al. (2007) in Figure~\ref{fig:alpha}.
For the majority of levels the change is $\lesssim1\%$.
There are two clear exceptions, corresponding to $3d\,^1\!D$ and $3d\,^3\!D$.
The cause is a programming error in our earlier renormalisation of Peach (1967) cross-sections.
There are only two levels affected.
Unfortunately, these are the upper levels of two of the most important lines
in primordial helium abundance works: $\lambda\lambda5876$ and $6678$.
The new results yield stronger emissivities for both lines.
The next largest differences in Figure~\ref{fig:alpha} correspond to levels with only weak optical lines.

\begin{figure}
\protect\resizebox{\columnwidth}{!}
{\includegraphics{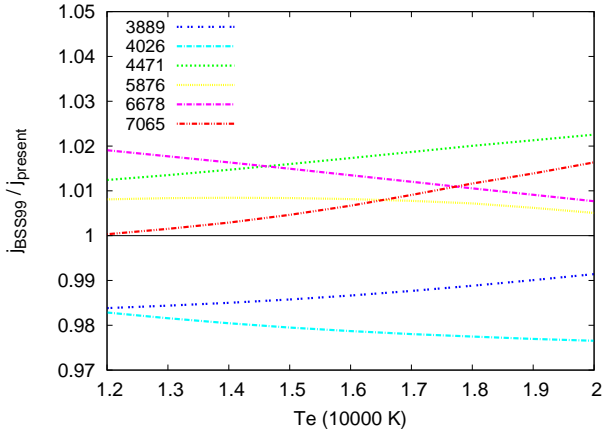}}
\caption{Ratio of BSS99 and present emissivities for several strong lines as a function of temperature with $n_e = 100$~cm$^{-3}$.  This figure should be compared with figure 5 of Aver, Olive, \& Skillman (2010).  The principal effect is that for both $\lambda\lambda 5876$ and $6678$, the present results are greater than the earlier Porter et al. works and are roughly the average of the BSS99 results and our earlier work.
Our $\lambda\lambda 3889$ and $7065$ results have also changed considerably, 
and (within the plotted temperature range) the ratios no longer cross the ``zero deviation line'' discussed by Aver, Olive, \& Skillman (2010).}
\label{fig:BSS99}
\end{figure}

We compare our new emissivities with BSS99 for several strong lines in Figure~\ref{fig:BSS99}.
This figure is directly comparable to figure 5 of Aver, Olive, \& Skillman (2010).
Consistent with the discussion above, $j_{5876}$ and $j_{6678}$ are now in better agreement with BSS99,
though important differences clearly remain.

Our calculations cover $5000 \le T \le 25000$~K (in $1000$~K steps) and $10^1 \le n_e \le 10^{14}$~cm$^{-3}$ (in 1 dex steps).
In addition to the lines published in our prior work, we have also included several weak infrared lines.
Table~\ref{table:lineID} lists the wavelength and upper and lower level designations of all reported lines.
Table~\ref{table:emis} contains a small subset of the full results that is most applicable to primordial helium abundance calculations.

\begin{table}
\centering
\caption{Wavelengths and upper and lower levels of reported lines.}
\begin{tabular}{rcc}
\hline
 Wavelength (\AA, Air)	&	Upper level	& Lower level	\\
\hline
2945		&	$5p\,{}^{3}\!P$	&	$2s\,{}^{3}\!S$	\\
3188		&	$4p\,{}^{3}\!P$	&	$2s\,{}^{3}\!S$	\\
3614		&	$5p\,{}^{1}\!P$	&	$2s\,{}^{1}\!S$	\\
3889		&	$3p\,{}^{3}\!P$	&	$2s\,{}^{3}\!S$	\\
3965		&	$4p\,{}^{1}\!P$	&	$2s\,{}^{1}\!S$	\\
4026		&	$5d\,{}^{3}\!D$	&	$2p\,{}^{3}\!P$	\\
4121		&	$5s\,{}^{3}\!S$	&	$2p\,{}^{3}\!P$	\\
4388		&	$5d\,{}^{1}\!D$	&	$2p\,{}^{1}\!P$	\\
4438		&	$5s\,{}^{1}\!S$	&	$2p\,{}^{1}\!P$	\\
4471		&	$4d\,{}^{3}\!D$	&	$2p\,{}^{3}\!P$	\\
4713		&	$4s\,{}^{3}\!S$	&	$2p\,{}^{3}\!P$	\\
4922		&	$4d\,{}^{1}\!D$	&	$2p\,{}^{1}\!P$	\\
5016		&	$3p\,{}^{1}\!P$	&	$2s\,{}^{1}\!S$	\\
5048		&	$4s\,{}^{1}\!S$	&	$2p\,{}^{1}\!P$	\\
5876		&	$3d\,{}^{3}\!D$	&	$2p\,{}^{3}\!P$	\\
6678		&	$3d\,{}^{1}\!D$	&	$2p\,{}^{1}\!P$	\\
7065		&	$3s\,{}^{3}\!S$	&	$2p\,{}^{3}\!P$	\\
7281		&	$3s\,{}^{1}\!S$	&	$2p\,{}^{1}\!P$	\\
9464		&	$5p\,{}^{3}\!P$	&	$3s\,{}^{3}\!S$	\\
10830	&	$2p\,{}^{3}\!P$	&	$2s\,{}^{3}\!S$	\\
11013 	&	$5p\,{}^{1}\!P$	&	$3s\,{}^{1}\!S$	\\
11969	&	$5d\,{}^{3}\!D$	&	$3p\,{}^{3}\!P$	\\
12527	&	$4p\,{}^{3}\!P$	&	$3s\,{}^{3}\!S$	\\
12756	&	$5p\,{}^{1}\!P$	&	$3d\,{}^{1}\!D$	\\ 
12785	&	$5f\,{}^{3}\!F$	&	$3d\,{}^{3}\!D$	\\ 
12790	&	$5f\,{}^{1}\!F$	&	$3d\,{}^{1}\!D$	\\
12846	&	$5s\,{}^{3}\!S$	&	$3p\,{}^{3}\!P$	\\
12968	&	$5d\,{}^{1}\!D$	&	$3p\,{}^{1}\!P$	\\
12985	&	$5p\,{}^{3}\!P$	&	$3d\,{}^{3}\!D$	\\ 
13412	&	$5s\,{}^{1}\!S$	&	$3p\,{}^{1}\!P$	\\
15084	&	$4p\,{}^{1}\!P$	&	$3s\,{}^{1}\!S$	\\
17003	&	$4d\,{}^{3}\!D$	&	$3p\,{}^{3}\!P$	\\ 
18556	&	$4p\,{}^{1}\!P$	&	$3d\,{}^{1}\!D$	\\
18685	&	$4f\,{}^{3}\!F$	&	$3d\,{}^{3}\!D$	\\
18697	&	$4f\,{}^{1}\!F$	&	$3d\,{}^{1}\!D$	\\
19089	&	$4d\,{}^{1}\!D$	&	$3p\,{}^{1}\!P$	\\
19543	&	$4p\,{}^{3}\!P$	&	$3d\,{}^{3}\!D$	\\
20427	&	$6p\,{}^{3}\!P$	&	$4s\,{}^{3}\!S$	\\
20581	&	$2p\,{}^{1}\!P$	&	$2s\,{}^{1}\!S$	\\
20602	&	$7d\,{}^{3}\!D$	&	$4p\,{}^{3}\!P$	\\ 
21118	&	$4s\,{}^{3}\!S$	&	$3p\,{}^{3}\!P$	\\
21130	&	$4s\,{}^{1}\!S$	&	$3p\,{}^{1}\!P$	\\
21608	&	$7f\,{}^{3}\!F$	&	$4d\,{}^{3}\!D$	\\ 
21617	&	$7f\,{}^{1}\!F$	&	$4d\,{}^{1}\!D$	\\ 
\hline
\end{tabular}
\label{table:lineID}
\end{table}

\begin{figure}
\protect\resizebox{\columnwidth}{!}
{\includegraphics{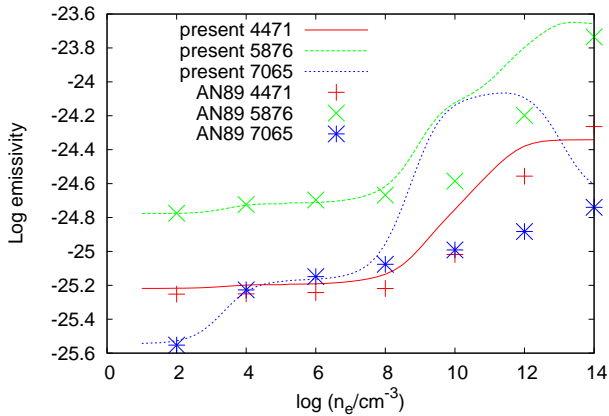}}
\caption{Comparison of present emissivity calculations to those by AN89 as a function of density at $10,000$K.}
\label{fig:AN89}
\end{figure}

Almog \& Netzer (1989, hereafter AN89) also presented He~I emissivities for electron densities up to $10^{14}$~cm$^{-3}$.
In Figure~\ref{fig:AN89} we compare present emissivities in $\lambda\lambda 4471$, $5876$ and $7065$ 
to the AN89 results (with $\tau_{3889} = 0$).
The calculations generally agree for $n_e \lesssim 10^{8}$~cm$^{-3}$ (where collisions do not dominate over radiative processes) 
and for $n_e \sim 10^{14}$~cm$^{-3}$ (where collisions are so dominant that levels are approaching
local thermodynamic equilibrium).
Critical densities for levels with $n\sim3$ or $4$ fall between these regions.
Collisions are dominated by excitations from the metastable $2s\,^3\!S$,
which is treated in both the present work and AN89.
Our work clearly has enhanced collisional excitation relative to AN89,
and differences between the two calculations are strongly correlated with the relative collisional contributions
given in table 5 of Porter et al. (2007).
Note, however, that those collisional contributions should not be added to the present tabulations;
they are already included.

A recalculation of the Monte Carlo error analyses performed in Porter et al. (2009) is beyond the scope of this paper.
We expect that our analysis of the relatively high-density `Galactic' model would be largely unchanged,
as collisional uncertainties should dominate both before and after the present work.
In the low-density `Extragalactic' model, however,
our `optimistic' uncertainties now seem somewhat more \emph{realistic}.
Total uncertainties are largely due to uncertainties in threshold photoionisation cross-sections.
One measure of those uncertainties (used by Porter et al. 2009)
is the difference ($\lesssim0.7\%$) between the extrapolated threshold cross-sections and the HS98 \textit{ab initio} results.
Goodness of the fits used to extrapolate threshold cross-sections,
including sensitivity of the fits to the number of terms used in the fitting procedure,
is another measure, and we obtain $\lesssim0.2\%$ for all levels with $n\le5$.

\begin{table*}
\centering
\caption{Emissivities of several He~I lines at conditions important for primordial abundance analyses.
This table is a small subset of the full results.  
Values are $4\pi j/n_e n_{\mathrm{He+}}$ in units $10^{-25}$~erg~cm$^{3}$~s$^{-1}$.}
\begin{tabular}{rrrrrrrr}
\hline
$T_e$~(K)& $n_e$ (cm$^{-3}$)&	 3889\AA	& 	4026\AA	&	 4471\AA	&	 5876\AA	&	 6678\AA	&	 7065\AA	\\
\hline
10000 & 10 	& 1.3889 & 0.2902 & 0.6098 & 1.6782 & 0.4706 & 0.2875 \\ 
11000 & 10 	& 1.2980 & 0.2652 & 0.5549 & 1.5112 & 0.4229 & 0.2727 \\ 
12000 & 10 	& 1.2194 & 0.2440 & 0.5086 & 1.3721 & 0.3833 & 0.2600 \\ 
13000 & 10 	& 1.1507 & 0.2257 & 0.4690 & 1.2546 & 0.3498 & 0.2487 \\ 
14000 & 10 	& 1.0901 & 0.2098 & 0.4347 & 1.1542 & 0.3213 & 0.2388 \\ 
15000 & 10 	& 1.0360 & 0.1959 & 0.4047 & 1.0674 & 0.2966 & 0.2299 \\ 
16000 & 10 	& 0.9875 & 0.1835 & 0.3783 & 0.9917 & 0.2751 & 0.2218 \\ 
17000 & 10 	& 0.9437 & 0.1726 & 0.3549 & 0.9252 & 0.2562 & 0.2145 \\ 
18000 & 10 	& 0.9039 & 0.1627 & 0.3340 & 0.8663 & 0.2395 & 0.2078 \\ 
19000 & 10 	& 0.8676 & 0.1539 & 0.3153 & 0.8138 & 0.2247 & 0.2017 \\ 
20000 & 10 	& 0.8343 & 0.1459 & 0.2983 & 0.7668 & 0.2113 & 0.1960 \\ 
10000 & 100 	& 1.3989 & 0.2904 & 0.6101 & 1.6768 & 0.4692 & 0.2975 \\ 
11000 & 100 	& 1.3100 & 0.2655 & 0.5557 & 1.5138 & 0.4224 & 0.2848 \\ 
12000 & 100 	& 1.2334 & 0.2444 & 0.5099 & 1.3787 & 0.3836 & 0.2738 \\ 
13000 & 100 	& 1.1666 & 0.2263 & 0.4708 & 1.2650 & 0.3509 & 0.2642 \\ 
14000 & 100 	& 1.1077 & 0.2105 & 0.4370 & 1.1682 & 0.3230 & 0.2557 \\ 
15000 & 100 	& 1.0553 & 0.1968 & 0.4075 & 1.0848 & 0.2990 & 0.2480 \\ 
16000 & 100 	& 1.0082 & 0.1846 & 0.3817 & 1.0123 & 0.2782 & 0.2409 \\ 
17000 & 100 	& 0.9657 & 0.1738 & 0.3587 & 0.9487 & 0.2598 & 0.2345 \\ 
18000 & 100 	& 0.9270 & 0.1641 & 0.3383 & 0.8926 & 0.2437 & 0.2285 \\ 
19000 & 100 	& 0.8917 & 0.1554 & 0.3200 & 0.8428 & 0.2293 & 0.2231 \\ 
20000 & 100 	& 0.8592 & 0.1475 & 0.3036 & 0.7984 & 0.2164 & 0.2181 \\ 
10000 & 1000 	& 1.4701 & 0.2928 & 0.6175 & 1.7310 & 0.4780 & 0.3754 \\ 
11000 & 1000 	& 1.3969 & 0.2687 & 0.5661 & 1.5893 & 0.4350 & 0.3771 \\ 
12000 & 1000 	& 1.3353 & 0.2486 & 0.5235 & 1.4761 & 0.4000 & 0.3790 \\ 
13000 & 1000 	& 1.2823 & 0.2315 & 0.4877 & 1.3842 & 0.3712 & 0.3807 \\ 
14000 & 1000 	& 1.2359 & 0.2168 & 0.4574 & 1.3084 & 0.3470 & 0.3816 \\ 
15000 & 1000 	& 1.1945 & 0.2040 & 0.4313 & 1.2449 & 0.3266 & 0.3818 \\ 
16000 & 1000 	& 1.1571 & 0.1929 & 0.4087 & 1.1910 & 0.3090 & 0.3811 \\ 
17000 & 1000 	& 1.1227 & 0.1830 & 0.3888 & 1.1444 & 0.2937 & 0.3797 \\ 
18000 & 1000 	& 1.0910 & 0.1743 & 0.3714 & 1.1041 & 0.2803 & 0.3778 \\ 
19000 & 1000 	& 1.0618 & 0.1666 & 0.3561 & 1.0706 & 0.2683 & 0.3765 \\ 
20000 & 1000 	& 1.0344 & 0.1596 & 0.3424 & 1.0408 & 0.2577 & 0.3746 \\ 
\hline
\end{tabular}
\label{table:emis}
\end{table*}

\section{Acknowledgments}

RLP thanks Leticia Mart\'{\i}n Hern\'andez for the suggestion to publish results for several of the weaker $\sim2\mu$m lines
(included in the full table), Keith MacAdam for helpful comments, and the referee for a valuable and practical suggestion.
The authors thank the University of Kentucky Center of Computational Sciences for a generous allotment of computer time.
GJF acknowledges support by NSF (0908877; 1108928; and 1109061), 
NASA (07-ATFP07-0124, 10-ATP10-0053, and 10-ADAP10-0073), JPL (RSA No 1430426), 
and STScI (HST-AR-12125.01, GO-12560, and HST-GO-12309).

\bsp

\label{lastpage}
\clearpage
\end{document}